\documentclass[aps,prb,a4paper,preprint,showpacs]{revtex4}
\usepackage{amsmath}
\usepackage{amsfonts}
\usepackage{amssymb}
\usepackage{slashbox}
\usepackage{graphicx}
\usepackage{amsbsy}

\begin{document}
\title{Conductance and shot noise in the helimagnet tunnel junction }
\author{Rui Zhu\renewcommand{\thefootnote}{*}\footnote{Corresponding author.
Electronic address:
rzhu@scut.edu.cn} and Guo-Yi Zhu}
\address{Department of Physics, South China University of Technology,
Guangzhou 510641, People's Republic of China }

\begin{abstract}

As a result of sinusoidal spatial modulation, helimagnet induces spin-dependent diffracted transmission. In this work, we propose a general scattering matrix treatment to the transport properties in helimagnets with arbitrary helical structures. Multiferroic properties can be considered by taking electric polarization and magnetic helix together into account. A monolayer magnetic helix can be treated as a $\delta$-barrier by diffraction theory. The conductance and shot noise properties of the normal metal/helimagnet/normal metal toy model are investigated. It is found that the shot noise is suppressed and demonstrates rise-and-fall variations as a result of interference between different diffracted channels. Sharp change of the shot noise occurs when one and both diffracted beams disappear into evanescent modes at the helimagnet spiral wave vector equal to one and two times the electron Fermi wave vector $q=k_F$ and $2 k_F$. We also considered the conductance and shot noise properties in a real multiferroic helimagnet $\rm{TbMnO_3}$ with ferroelectric and helimagnetism coexistent. It is found that clockwise and counterclockwise spin helix is distinctly separated in the conductance and shot noise spectrum. Their variation pattern is a combined result of helimagnetism and electric polarization.

\end{abstract}


\maketitle

\narrowtext

\section{Introduction}

Diffraction is centuries-old understanding of the single-beam-in multi-beam-out phenomenon in optics, acoustics, quantum mechanics, and all wave equation governed scattering processes, when the middle media has some property periodically varying in space. Naturally we should know that as a simple case transmission of the electron through a sinusoidal-height quantum potential barrier or sinusoidal-depth well demonstrates diffraction effect. Although in vast numerical treatments such as the plane-wave expansion method for metamaterials, this problem is just a building block within, we think it is important to particularly treat the quantum grating effect by developing a general scattering method and provide the detailed physical picture \emph{ab initio} from the Schr\"odinger equation, which to our knowledge is not covered in literature. Also, the sinusoidal-height quantum potential barrier itself can be the model of a real material and a real transport device based on it. The recent widely-focused material, helimagnet (HM), lends a very appropriate platform.

The HM is a kind of magnetic state\cite{Ref13} with its spin spiraling in two or three dimensions characterized by a single spiral wavevector ${\bf{Q}}$, which is different from conventional spatially uniform ferromagnet and antiferromagnet.
 When a single electron passes through the HM structure, spin-dependent diffraction occurs. Recently the transport properties of the HM-embedded devices are targeted from different view angles in literature. Manchon \emph{et al.}\cite{Ref42} and us\cite{Ref43} observed the spin-dependent diffraction effect in the transmission at the ferromagnet/HM interface and through the thin-layer HM junction, respectively. Some functional devices were proposed based on the HM such as the persistent spin currents\cite{Ref19}, spin-field-effect transistor\cite{Ref17}, tunneling anisotropic magnetoresistance\cite{Ref18}, and spin resonance\cite{Ref13}. Conductance characteristic of the HM spin configuration and spiral period was found in the Fano resonance spectrum of a quasi-one-dimensional WG containing a thin conducting HM layer as a donor impurity\cite{Ref52}. Ac gate potential driven Quantum pumping behavior was also investigated recently\cite{ZhuPLA2014}. Although some works were done to investigate the HM-related transport properties, there lacks an overall description of the diffraction scenario.

As far as a good transport approach can go, beyond the conductance, we also considered the diffraction governed shot noise properties. As a consequence of the
quantization of charge and defined by quantum contribution in the current fluctuations, shot noise is useful to obtain information
on a system which is not available through conductance
measurements\cite{BlanterPR2000}. Two of the most significant shot noise experiments are  carrier charge confirmations of the Cooper pair\cite{CooperShotNoise} and Laughlin quasiparticle\cite{LaughlinShotNoise}. In most cases, the properties of quantum correlation are reflected in the Fano factor $F$, which is defined by proportion of the real shot noise $S$ to Poisson noise $2eI$ ($I$ is the average current), the latter of which corresponds to single quasiparticle transmission without correlation. Therefore, some levels of the Fano factor have typical physical meaning. $F=1$ characterizes Poisson noise. Besides the ideal case, the Fano factor approaches $1$ when the transmission is extremely small corresponding to uncorrelated transport and closed channel in ballistic tunneling. $F=0$ characterizes full correlation and maximal quantum coherence. In real conductors, the Fano factor approaches $0$ when the transmission reaches $1$ corresponding to open channels in ballistic transport. In some cases with strong electron-electron interaction involved\cite{EnhacedShotNoise}, the shot noise can be enhanced beyond $1$. $F=1/2$ characterizes the effect of Pauli exclusion and $F=1/3$ characterizes diffusive transport when open and closed channels distributes in disorder such as diffusive metals\cite{DiffusiveMetal} and graphene\cite{Tworzydlo}. With understanding of the physics underlying different Fano factor levels, we could suppose that diffraction enhances transmission, different diffraction channels have strong coherence, and the shot noise should be thus suppressed. Our theory would confirm this supposition in detail.

Also, the transport properties are governed by diffraction. For the HM spiral wave vector $q$ larger than two times the electron Fermi wave vector $2 k_F$, both diffraction beams degrade into evanescent surface modes and do not contribute to the transmission, in which case the transmission is identical to that of a plain barrier. As a result, sharp change in the conductance, shot noise, and Fano factor occurs at $q=2 k_F$, lending a potential transport measurement of $q$.

\section{Spin Spiral Monolayer Toy Model}

\subsection{Theoretic Formulism}

Our model is sketched in Fig. 1 (a). An HM interlayer is put between two semiinfinite free regions extending in the $x$-$y$ plane. For electrons, those free regions can be normal metal leads. Transport direction is along the $z$ coordinate. The HM spin varies in space with the vector field
\begin{equation}
{{\bf{n}}_r} = \left[ {\sin \left( {qx} \right),0,\cos \left( {qx} \right)} \right].
\label{SpinVectorFiled}
\end{equation}
$q$ is the spin wave vector and the helix is two dimensional modulating in the $x$ direction. The spin exchange between the free electron and the HM magnetization giving rise to a space-dependent Zeeman term in the Hamiltonian, which is
\begin{equation}
H =  - \frac{{{\hbar ^2}}}{{2{m^*}}}{\nabla ^2} + \left( {J{{\bf{n}}_r} \cdot {\bf{\sigma }} + {V_0}} \right)\delta \left( z \right).
\label{HMHamiltonian}
\end{equation}
Here, $m^*$ is the HM electron effective mass. $J$ refers to
space and momentum averages of the exchange coupling strength. $\bf{\sigma }$ is the Pauli matrix. We assume an ultrathin HM layer located at the $z=0$ plane, so its effect in the Hamiltonian can be approximated by a $\delta$-function. $V_0$ is the electrostatic potential of the HM. For insulating HM, $V_0 >0$ is a barrier potential; for conducting HM, $V_0 <0$ is a well potential. We consider the former case.

Scattered by the HM interlayer, the electron wave function with incidence, reflection, and transmission beams in the two free regions can be written as
\begin{equation}
\left\{ \begin{array}{l}
{\psi _I  (x,z)} = \sum\limits_{n =  - \infty ,\sigma }^{ + \infty } {\left( {A_{n\sigma }^i{e^{i{k_{xn}}x}}{e^{i{k_{zn}}z}}{\chi _\sigma } + A_{n\sigma }^o{e^{i{k_{xn}}x}}{e^{ - i{k_{zn}}z}}{\chi _\sigma }} \right)} ,\begin{array}{*{20}{c}}
{}&{z < 0,}
\end{array}\\
{\psi _{II} (x,z) } = \sum\limits_{n =  - \infty ,\sigma }^{ + \infty } {\left( {B_{n\sigma }^i{e^{i{k_{xn}}x}}{e^{ - i{k_{zn}}z}}{\chi _\sigma } + B_{n\sigma }^o{e^{i{k_{xn}}x}}{e^{i{k_{zn}}z}}{\chi _\sigma }} \right)} ,\begin{array}{*{20}{c}}
{}&{z > 0,}
\end{array}
\end{array} \right.
 \label{WeveFunctions}
\end{equation}
where $k_{xn}=k_x + nq$ and ${k_{zn}} = \sqrt {k_F^2 - k_y^2 - k_{xn}^2} $ with the Fermi wave vector ${k_F} = {{\sqrt {2{m_e}{E_F}} } \mathord{\left/
 {\vphantom {{\sqrt {2{m_e}{E_F}} } \hbar }} \right.
 \kern-\nulldelimiterspace} \hbar }$, $E_F$ the electron Fermi energy and $m_e$ the free electron mass. $A_{n\sigma }^i$ and $B_{n\sigma }^i$ are the probability amplitudes of the incoming waves from the lower and upper leads, respectively, while $A_{n\sigma }^o$ and $B_{n\sigma }^o$ are those of the outgoing waves. Diffraction occurs in the $x$ direction. Translation symmetry protects the plane wave component in the $y$-direction $e^{ik_y y}$ unchanged during transmission.

By continuity equation at the HM interface
\begin{equation}
{\psi _I}\left( {x,{0^ - }} \right) = {\psi _{II}}\left( {x,{0^ + }} \right),
\end{equation}
and
\begin{equation}
\frac{{{\hbar ^2}}}{{2{m_e}}}{\left. {\frac{{\partial {\psi _I}}}{{\partial z}}} \right|_{z = {0^ - }}} + \left( {{V_0} + { J \bf{w}}} \right){\psi _I}\left( {x,{0^ - }} \right) = \frac{{{\hbar ^2}}}{{2{m_e}}}{\left. {\frac{{\partial {\psi _{II}}}}{{\partial z}}} \right|_{z = {0^ + }}},
\end{equation}
with
\begin{equation}
{\bf{w}} = \left[ {\begin{array}{*{20}{c}}
{\cos \left( {qx} \right)}&{\sin \left( {qx} \right)}\\
{\sin \left( {qx} \right)}&{ - \cos \left( {qx} \right)}
\end{array}} \right],
\end{equation}
the scattering matrix relation
\begin{equation}
\left( {\begin{array}{*{20}{c}}
{A_{n\sigma }^o}\\
{B_{n\sigma }^o}
\end{array}} \right) =\sum _m { \left( {\begin{array}{*{20}{c}}
{r_{nm}^{\sigma \tau }} & {{t'}_{nm}^{\sigma \tau }}\\
{t_{nm}^{\sigma \tau }} & {{r'}_{nm}^{\sigma \tau }}
\end{array}} \right)\left( {\begin{array}{*{20}{c}}
{A_{m\tau }^i}\\
{B_{m\tau }^i}
\end{array}} \right)}=\sum _m {{{\bf{\hat s}}_{\sigma \tau }}\left( {{{\bf{k}}_n},{{\bf{k}}_m}} \right)\left( {\begin{array}{*{20}{c}}
{A_{m\tau }^i}\\
{B_{m\tau }^i}
\end{array}} \right)}
\end{equation}
can be obtained. $r_{nm}^{\sigma \tau }$ and $t_{nm}^{\sigma \tau }$ represent respectively the reflection and transmission amplitudes from the spin-$\tau$, $m$th-order diffraction channel to the spin-$\sigma$, $n$th-order diffraction channel. ${r'}_{nm}^{\sigma \tau }$ and ${t'}_{nm}^{\sigma \tau }$ are the corresponding backward amplitudes. ${\bf{k}}_n=(k_{xn},k_y,k_{zn})$ is the three-dimensional wavevector labeling diffraction channels. Considering the real current flux, the scattering matrix
\begin{equation}
{{\bf{\hat S}}_{\sigma \tau }}\left( {{{\bf{k}}_n},{{\bf{k}}_m}} \right) = \sqrt {\frac{{{\mathop{\rm Re}\nolimits} \left( {{k_{zn}}} \right)}}{{{\mathop{\rm Re}\nolimits} \left( {{k_{zm}}} \right)}}} {{\bf{\hat s}}_{\sigma \tau }}\left( {{{\bf{k}}_n},{{\bf{k}}_m}} \right) .
\label{DefiningBigScatteringMatrix}
\end{equation}

With the scattering matrix ${{\bf{\hat S}}_{\sigma \tau }}\left( {{k_n},{k_m}} \right)$, the conductance and shot noise at low temperatures can be calculated as follows\cite{BlanterPR2000}.
\begin{equation}
G = \frac{e^2}{{{h}}}\int_0^{2\pi } {\int_0^{{\pi  \mathord{\left/
 {\vphantom {\pi  2}} \right.
 \kern-\nulldelimiterspace} 2}} {\sum\limits_{\sigma ,\tau ,n} {{{\left| {{\bf{\hat S}}_{\sigma \tau }^{RL}\left( {{k_n},{k_0}} \right)} \right|}^2}} k_F^2\sin {\theta _{in}}d{\theta _{in}}d{\phi _{in}}} } .
 \label{G}
\end{equation}
\begin{equation}
S = \frac{{2{e^3}}}{h}\int_0^{2\pi } {\int_0^{{\pi  \mathord{\left/
 {\vphantom {\pi  2}} \right.
 \kern-\nulldelimiterspace} 2}} {\sum\limits_{\sigma ,\tau ,n} {\left\{ {{{\left| {{\bf{\hat S}}_{\sigma \tau }^{RL}\left( {{k_n},{k_0}} \right)} \right|}^2}\left[ {1 - {{\left| {{\bf{\hat S}}_{\sigma \tau }^{RL}\left( {{k_n},{k_0}} \right)} \right|}^2}} \right]} \right\}} k_F^2\sin {\theta _{in}}d{\theta _{in}}d{\phi _{in}}} } .
 \label{S}
\end{equation}
Here, $\theta _{in}$ and $\phi _{in}$ are the incident polar and azimuthal angles, respectively. The Fano factor can be defined by $F=S/(2eG)$.

\subsection{Numerical Results and Interpretations}

Numerical results of the diffracted transmission for $q=0.5 k_F$ are shown in Fig. 1. $T_{0,\pm 1}$ is defined as $\sum\limits_{\sigma ,\tau } {{{\left| {{\bf{\hat S}}_{\sigma \tau }^{RL}\left( {{k_{0, \pm 1}},{k_0}} \right)} \right|}^2}} $. The zero-order transmission $T_0$ has spherical symmetry in the incident angle space, namely, in the electron wave vector space. It approaches maximum at normal incidence, which is natural for standard barrier tunneling as the wave vector in the propagating direction and hence the current flux approaches maximum. The $1$ and $-1$ order transmission is at minimum in normal incidence due to the $x$-$z$ plane symmetry of the HM spiral. When the incident angle increases, $T_1$ and $T_{-1}$ increases, approaches maximum and abruptly disappears into evanescent modes for the incident azimuthal angle $\phi _{in}=0$ and polar angle $\theta _{in} =\pi /6$ and $\theta _{in} =- \pi /6$, respectively. The angle can be analytically obtained by the equation of ${k_{z, \pm 1}} = \sqrt {k_F^2 - {{\left( {{k_F}\sin {\theta _{in}}\sin {\phi _{in}}} \right)}^2} - {{\left( {{k_F}\sin {\theta _{in}}\cos {\phi _{in}} \pm 0.5{k_F}} \right)}^2}} =0$. To the other side of the incident sphere, $T_1$ and $T_{-1}$ disappears at the glazing angle.

To investigate the conductance and shot noise properties of the HM tunnel junction, angle averaged quantities of Eqs. (\ref{G}) and (\ref{S}) are shown in Fig. 2. From panel (a), it can be seen that the conductance monotonously increases with the Fermi wave vector $k_F$. It weakly depends on the HM spiral wave vector $q$ and is not visible in the figure. The conductance increase in $k_F$ is due to larger current flux and more contributing channels. The weak dependence on $q$ is due to diffusion of the diffraction effect by angle average. The angle-averaged shot noise as functions of $q$ and $k_F$ is shown in panel (b). Prominent diffraction effect can be seen in the noise spectrum. As a result of diffraction, different diffracted channels interact and give rise to the multiple peaks in the shot noise. The pink and red dotted lines correspond to $q= k_F$ and $q=2 k_F$, respectively. For $q$ larger than $2 k_F$, all diffracted waves degrade into evanescent modes and variation of the shot noise became smooth without abrupt rises and falls. For $q$ larger than $k_F$ and smaller than $2 k_F$, $T_1$ disappears for all positive incident angles and $T_{-1}$ disappears for all negative incident angles. For $q$ smaller than $k_F$, diffracted transmission from part of the incident semi-sphere becomes evanescent and does not contribute to the transport as can be seen in Fig. 1. At the two sides of the pink dotted line in Fig. 2 (b) when $q$ is close to $k_F$, strength of the two diffracted waves $T_1$ and $T_{-1}$ matches each other, giving rise to maximal channel-coherence. Therefore, peaks of the shot noise are dramatically enhanced at the two side of the pink dotted line labeling $q= k_F$. When the absolute value of $k_F$ is small relative to $q$, almost no diffraction channel exists and the shot noise is extremely small. There is an abrupt increase in the shot noise when diffracted channels begin to contribute to the transport.

The relative strength of the shot noise in comparison with the Poisson noise is measured by the Fano factor. In Fig. 2 (c) and (d), we show numerical results of the Fano factor. From panel (c), it can be seen that the absolute value of the Fano factor is smaller than $1/3$ throughout the considered parameter space. As summarized in the Introduction, this regime is between the complete ballistic tunneling regime of $F=0$ and the diffusive tunneling regime of $F=1/3$. Our considered HM tunnel junction is a low $\delta$-barrier with the spiral modulating in the spin space. The $\delta$-barrier strength $V_0=50$ meV$ \cdot $\AA and the HM exchange coupling strength $J=20$ meV$ \cdot $\AA. Therefore, even without the spiral modulation, the transmission is very large and close unity. The spiral modulation trifurcated the incident electron plane wave into the $0$ and $\pm 1$ order diffracted waves. The main $0$-order transmission still governs with $T_0$ approaching $1$ and $T_{\pm 1}$ three to four orders smaller than it. These transmission properties can be clearly seen in Fig. 1. As an effect, the system approximates an open conductor and the Fano factor is very small. However, coherence between different diffraction channels significantly influences the shot noise. There are two prominent phenomena. One is that the shot noise is dramatically suppressed relative to the poisson value. The other is that the shot noise demonstrates rise-and-fall variations. Therefore, the Fano factor is smaller for larger $k_F$ for stronger transmission and more contributing channels. Also oscillations can be seen in the Fano factor specified in the panel (d). The Fano factor oscillates as a function of $q$ for $q<2 k_F$. The oscillation is small in comparison with its decrease as a function of $k_F$ and is not prominently seen in the panel (c). When $q> 2 k_F$ diffraction disappears, the shot noise properties resemble a plain barrier. For small Fermi energies, there is a dividing line in the shot noise between the diffraction affected transport for $q< 2 k_F$ and the ordinary barrier scattering governed transport for $q>2 k_F$, which is already seen in the panel (b). In the panel (d) the black and red star symbols label the dividing position of $q=2 k_F$ for different Fermi energies. The shot noise properties also lend a potential detection of the HM spiral period and spin wave vector.

\section{Investigation of the Multiferroic Helimagnet $\rm{TbMnO_3}$ }

Our theoretical treatment introduced in the previous section can be generalized into arbitrary spin spiral structures with finite thickness. In this section, the multiferroic HM\cite{YamasakiPRL2007} $\rm{TbMnO_3}$ is considered by taking into account the combined effect of its helimagnetism and electric polarization. The helimagnetic structure and electric potential profile are sketched in Fig. 3.

\subsection{Theoretic Formulism}

The spiral magnetic structure can be described as
\begin{equation}
{{\bf{M}}_i} = {{\bf{m}}_b}\cos \left( {2\pi {{\bf{Q}}_m} \cdot {{\bf{R}}_i}} \right) + {{\bf{m}}_c}\sin \left( {2\pi {{\bf{Q}}_m} \cdot {{\bf{R}}_i}} \right).
\label{HelimagnetismTbMnO3}
\end{equation}
It was reported\cite{YamasakiPRL2007} for $\rm{TbMnO_3}$ that ${\bf{m}}_b \approx (0,3.9,0)\mu _B$ and ${\bf{m}}_c \approx (0,0,2.8)\mu _B$ at 15 K. We define ${\bf{m}}_b = (0,m_b,0)\mu _B$ and ${\bf{m}}_c = (0,0,m_c)\mu _B$ to consider their variation with temperature. ${\bf{R}}_i$ extends in the three-dimensional crystal lattice. ${\bf{R}}_i$ and ${\bf{Q}}_m$ can be expressed relatively as ${{\bf{Q}}_m} = \left( {0, \pm 2\pi q/b, 2 \pi/c } \right)$ and ${{\bf{R}}_i} = \left( {{n_a} a,{n_b}b,{n_c}\frac{c}{2}} \right),$ with $n_a$, $n_b$, and $n_c$ arbitrary integers. From experimental results\cite{YamasakiPRL2007}, $q \approx 0.27$. It could be seen from Eq. (\ref{HelimagnetismTbMnO3}) that the magnetic moments are reversed every other $c/2$ layer (see Fig. 3).

Every atom layer in the transport direction ($c/z$) can be treated as a $\delta$-barrier. To consider a finite-thickness HM, we use a transfer matrix technique on multiple $\delta$-barriers. Magnetic spiral variation in the $b/y$-direction can be approximated to be continuous. Hamiltonian in each atomic layer is Eq. (\ref{HMHamiltonian}), with the exchange coupling terms in neighboring atom layers
\begin{equation}
J{{\bf{n}}_r} \cdot {\bf{\sigma }} = J\left[ {\begin{array}{*{20}{c}}
   {{ \tilde m_c}\sin \left( {qy} \right)} & { - i{\tilde m_b}\cos \left( {qy} \right)}  \\
   {i{\tilde m_b}\cos \left( {qy} \right)} & { - {\tilde m_c}\sin \left( {qy} \right)}  \\
\end{array}} \right],
\end{equation}
\begin{equation}
J{{\bf{n}}_r} \cdot {\bf{\sigma }} = J\left[ {\begin{array}{*{20}{c}}
   { - {\tilde m_c}\sin \left( {qy} \right)} & {i{\tilde m_b}\cos \left( {qy} \right)}  \\
   { - i{\tilde m_b}\cos \left( {qy} \right)} & {{\tilde m_c}\sin \left( {qy} \right)}  \\
\end{array}} \right],
\end{equation}
respectively. $\tilde m_b =m_b/ \sqrt {m_b^2+m_c^2}$ and $\tilde m_c =m_c/ \sqrt {m_b^2+m_c^2}$ are altered to normalize ${\bf{n}}_r$.

Ferroelectric polarization and helimagnetism coexist and are related in the multiferroic HM $\rm{TbMnO_3}$. Dielectric constant is reported\cite{Ref18} to be $\varepsilon  = 30$. As a result of the screening charges, the potential difference in the $c/z$-direction between the two interfaces of the HM could be approximated to be $\Delta U=P_c d/(\varepsilon_0  \varepsilon)$, with $P_c$ $c$-component of the electric polarization (it has only $c$-component) and $d$ thickness of the HM. Experimental data of $P_c$ from Ref. \onlinecite{YamasakiPRL2007} are used.

To consider scattering through an arbitrary atom layer approximated into a $\delta$-barrier, the flux-normalized wave functions at the two interfaces could be written as
\begin{equation}
\left\{ {\begin{array}{*{20}{c}}
   {{\psi _I (y,z)} = \sum\limits_{n,\sigma } {\left( {\frac{{A_{n\sigma }^i}}{{\sqrt {{k_n}} }}{e^{i{q_n}y + i{k_n}z}}{\chi _\sigma } + \frac{{A_{n\sigma }^o}}{{\sqrt {{k_n}} }}{e^{i{q_n}y - i{k_n}z}}{\chi _\sigma }} \right)} ,} \hfill & {z < 0,} \hfill  \\
   {{\psi _{II} (y,z)} = \sum\limits_{n,\sigma } {\left( {\frac{{B_{n\sigma }^i}}{{\sqrt {{k_n}} }}{e^{i{q_n}y - i{k_n}z}}{\chi _\sigma } + \frac{{B_{n\sigma }^o}}{{\sqrt {{k_n}} }}{e^{i{q_n}y + i{k_n}z}}{\chi _\sigma }} \right)} ,} \hfill & {z > 0.} \hfill  \\
\end{array}} \right.
\label{DiffractedWaveFunctions}
\end{equation}
Since the real position of a $\delta$-barrier does not affect its transmission, here we assume it locate at $z=0$. In Eq. (\ref{DiffractedWaveFunctions}), summation is over all diffraction and spin channels, ${q_n} = {k_y} + n\tilde q$, and ${k_n} = \sqrt {k_F^2 - k_x^2 - q_n^2} $ with $k_F$ the Fermi wave vector and $\tilde q =2 \pi q/b$. $k_x$ is conserved in transmission due to translational invariance in the $a/x$ crystal direction. $\chi _{\sigma}$ are arbitrary eigenspinors and we set them to be ${\chi _ \uparrow } = {\left( {\begin{array}{*{20}{c}}
   1 & 0  \\
\end{array}} \right)^{\rm{T}}}$ and ${\chi _ \downarrow } = {\left( {\begin{array}{*{20}{c}}
   0 & 1  \\
\end{array}} \right)^{\rm{T}}}$ for algebra simplicity. $A_{n\sigma }^i$($B_{n\sigma }^i$) and $A_{n\sigma }^o$($B_{n\sigma }^o$) are the incident and outgoing probability amplitudes at the $z<0$($z>0$) interfaces, respectively. A cutoff of ${n_{\max }} =8  > 2{k_F}/\tilde q$ secures accuracy in our numerical treatment.

By continuity at the $z=0$ HM interface
\begin{equation}
{\psi _I}\left( {y,{0^ - }} \right) = {\psi _{II}}\left( {y,{0^ + }} \right),
\end{equation}
and
\begin{equation}
\frac{{{\hbar ^2}}}{{2{m_e}}}{\left. {\frac{{\partial {\psi _I}}}{{\partial z}}} \right|_{z = {0^ - }}} + \left( {{V_0} + J{{\bf{n}}_r} \cdot {\bf{\sigma }}} \right){\psi _I}\left( {y,{0^ - }} \right) = \frac{{{\hbar ^2}}}{{2{m_e}}}{\left. {\frac{{\partial {\psi _{II}}}}{{\partial z}}} \right|_{z = {0^ + }}},
\end{equation}
we could obtain the following matrix element equations
\begin{equation}
A_{n\sigma }^i + A_{n\sigma }^o = B_{n\sigma }^i + B_{n\sigma }^o,
\label{ContinuityOne}
\end{equation}
and
\begin{equation}
\begin{array}{l}
 i\sqrt {{k_n}} \left( {B_{n\sigma }^o - B_{n\sigma }^i + A_{n\sigma }^o - A_{n\sigma }^i} \right) \\
  = \sum\limits_{m,\sigma '} {\left[ {\left( {\frac{{{V_0}{\delta _{nm}}{\delta _{\sigma \sigma '}}}}{{\sqrt {{k_n}} }} + \frac{{{J_{ - 1\sigma \sigma '}}{\delta _{n - 1,m}}}}{{\sqrt {{k_{n - 1}}} }} + \frac{{{J_{ + 1\sigma \sigma '}}{\delta _{n + 1,m}}}}{{\sqrt {{k_{n + 1}}} }}} \right)\left( {B_{m\sigma '}^i + B_{m\sigma '}^o} \right)} \right]} , \\
 \end{array}
 \label{ContinuityTwo}
\end{equation}
with matrices operating on the spin space
\begin{equation}
{J_{ - 1}} = J\left( {\begin{array}{*{20}{c}}
   { - \frac{{{{\tilde m}_c}}}{{2i}}} & {\frac{{i{{\tilde m}_b}}}{2}}  \\
   { - \frac{{i{{\tilde m}_b}}}{2}} & {\frac{{{{\tilde m}_c}}}{{2i}}}  \\
\end{array}} \right),\begin{array}{*{20}{c}}
   {} & {{J_{ + 1}} = J\left( {\begin{array}{*{20}{c}}
   {\frac{{{{\tilde m}_c}}}{{2i}}} & {\frac{{i{{\tilde m}_b}}}{2}}  \\
   { - \frac{{i{{\tilde m}_b}}}{2}} & { - \frac{{{{\tilde m}_c}}}{{2i}}}  \\
\end{array}} \right).}  \\
\end{array}
\label{JPL1}
\end{equation}

With Eqs. (\ref{ContinuityOne}) and (\ref{ContinuityTwo}), we could obtain the transfer matrix connecting the probability amplitudes on the two interfaces of the HM $\delta$-barrier as
\begin{equation}
\left( {\begin{array}{*{20}{c}}
   {{B^i}}  \\
   {{B^o}}  \\
\end{array}} \right) = \left( {\begin{array}{*{20}{c}}
   {{M_{ii}}} & {{M_{io}}}  \\
   {{M_{oi}}} & {{M_{oo}}}  \\
\end{array}} \right)\left( {\begin{array}{*{20}{c}}
   {{A^i}}  \\
   {{A^o}}  \\
\end{array}} \right),
\label{DefiningSingleLayerTransferMatrix}
\end{equation}
in which $A^{i/o}$ and $B^{i/o}$ are matrices made up of the precious elements and a Kronecker product between the diffraction and spin spaces.
The total transfer matrix of multiple $\delta$-barrier follows as
\begin{equation}
{M_f} = \prod\limits_{i = 1}^N {{M_i}}  = {M_N} \cdot  \ldots  \cdot {M_2} \cdot {M_1},
\end{equation}
with
\begin{equation}
{M_i} = \left( {\begin{array}{*{20}{c}}
   {{M_{ii}}} & {{M_{io}}}  \\
   {{M_{oi}}} & {{M_{oo}}}  \\
\end{array}} \right)
\end{equation}
for odd $i$'s. $M_i$ with even $i$'s can be obtained by changing $J$ into $-J$ of Eq. (\ref{JPL1}). Secured by the flux normalization of the eigen-spinor wave functions, the scattering matrix ${\bf{\hat S}}$ defined in Eq. (\ref{DefiningBigScatteringMatrix}) can be derived from the transfer matrix $M_f$ as
\begin{equation}
{\bf{\hat S}} = \left( {\begin{array}{*{20}{c}}
   {{M_{aa}}} & {{M_{ab}}}  \\
   {{M_{ab}}} & {{M_{bb}}}  \\
\end{array}} \right),
\end{equation}
with
\begin{equation}
\left\{ \begin{array}{l}
 {M_{aa}} =  - {\left( {{M_{fio}}} \right)^{ - 1}}{M_{fii}}, \\
 {M_{ab}} = {\left( {{M_{fio}}} \right)^{ - 1}}, \\
 {M_{ba}} = {M_{foi}} - {M_{foo}}{\left( {{M_{fio}}} \right)^{ - 1}}{M_{fii}}, \\
 {M_{bb}} = {M_{foo}}{\left( {{M_{fio}}} \right)^{ - 1}}. \\
 \end{array} \right.
\end{equation}
Here, relation (\ref{DefiningSingleLayerTransferMatrix}) also holds for $M_f$ connecting probability amplitudes of the beginning and ending interfaces of the total $N$-layer scattering. The total transmission probability can be defined as
$\sum\limits_{\sigma ,\tau ,n} {{{\left| {\hat S_{\sigma \tau }^{RL}\left( {{k_n},{k_0}} \right)} \right|}^2}} $.

To consider the helimagnetism variation in temperature, the conductance, bias current, and noise can be expressed as follows,
\begin{equation}
G = \frac{{{e^2}}}{h}\int_{ - \infty }^{ + \infty } {\int_0^{2\pi } {\int_0^{{\pi  \mathord{\left/
 {\vphantom {\pi  2}} \right.
 \kern-\nulldelimiterspace} 2}} {\sum\limits_{\sigma ,\tau ,n} {{{\left| {{\bf{\hat S}}_{\sigma \tau }^{RL}\left( {{k_n},{k_0}} \right)} \right|}^2}} \frac{{df\left( E \right)}}{{dE}}k_F^2\sin {\theta _{in}}d{\theta _{in}}d{\phi _{in}}} } dE} ,
\end{equation}
\begin{equation}
I = \frac{e}{h}\int_{ - \infty }^{ + \infty } {\int_0^{2\pi } {\int_0^{{\pi  \mathord{\left/
 {\vphantom {\pi  2}} \right.
 \kern-\nulldelimiterspace} 2}} {\sum\limits_{\sigma ,\tau ,n} {{{\left| {{\bf{\hat S}}_{\sigma \tau }^{RL}\left( {{k_n},{k_0}} \right)} \right|}^2}} \left[ {f\left( {E - eV} \right) - f\left( E \right)} \right]k_F^2\sin {\theta _{in}}d{\theta _{in}}d{\phi _{in}}} } dE} ,
\end{equation}
\begin{equation}
S = \frac{{{e^2}}}{{2\pi \hbar }}\sum\limits_{\gamma \delta } {\sum\limits_{mn} {\sum\limits_{{\sigma _1}{\sigma _2}} {\int {dEd{\theta _{in}}d{\phi _{in}}A_{\gamma \delta {\sigma _1}{\sigma _2}}^{mn}\left( {L;E} \right)A_{\delta \gamma {\sigma _2}{\sigma _1}}^{nm}\left( {L;E} \right)} } } } ,
\end{equation}
with
\begin{equation}
A_{\gamma \delta {\sigma _1}{\sigma _2}}^{mn}\left( {L;E} \right) = {\delta _{mn}}{\delta _{{\sigma _1}{\sigma _2}}}{\delta _{L\gamma }}{\delta _{L\delta }} - \sum\limits_{k,{\sigma _3}} {s_{\gamma L;km;{\sigma _3}{\sigma _1}}^\dag \left( E \right){s_{L\delta ;nk;{\sigma _2}{\sigma _3}}}\left( E \right)} .
\end{equation}
The the equilibrium, or Nyquist--Johnson noise is $S_t=4 k_B T G$.
Therefore the Fano factor is defined as $F=(S-S_t)/(2eI)$. To prominently see the quantum transport effect, we define the relative conductance $G_r$ as
\begin{equation}
{G_r} = \frac{G}{{\int_{ - \infty }^{ + \infty } {\frac{{df\left( E \right)}}{{dE}}dE} }}.
\label{RelativeConductance}
\end{equation}

\subsection{Numerical Results and Interpretations}

To obtain results comparable with the experiment, we used real parameters directly simulated from Ref. \onlinecite{YamasakiPRL2007}. The simulated ferroelectric polarization $P_c$ and spiral ellipticity $m_c / m_b$ as a function of the temperature are shown in Fig. 4. From Ref. \onlinecite{YamasakiPRL2007}, it can be seen that $m_b$ linearly varies with the temperature. $m_c$ is nonzero for temperatures below the HM transition temperature $T_C =27 $ K and has a power law relation to the temperature. By fitting to the experimental data, we assume
\begin{equation}
m_b=-0.14 T+6.07,
\end{equation}
and
\begin{equation}
{m_c} = \left\{ {\begin{array}{*{20}{c}}
   {\frac{{{{\left( {27 - T} \right)}^{{1 \mathord{\left/
 {\vphantom {1 2}} \right.
 \kern-\nulldelimiterspace} 2}}}}}{{{m_b}}},} \hfill & {T < {T_C}} \hfill  \\
   {0,} \hfill & {{T_C} < T < {T_N},} \hfill  \\
\end{array}} \right.
\end{equation}
and
\begin{equation}
P_c = A \sin (2\pi qb) m_b m_c,
\end{equation}
with the temperature $T$ measured in degree kelvin and $T_N = 42$ K is the long-range-spin-order-emerging temperature. $A=40$ $\mu $F is an experimentally determined constant. The ferroelectric polarization generates screening charges on the interfaces between the HM and the metal electrodes, which results in an electric field and electric voltage difference. We divide the voltage increase (decrease) into ultrathin steps with one layer corresponding to a real atomic layer illustrated in Fig. 3 (b) and the voltage change between neighboring atomic layers $\Delta U=P_c d/(\varepsilon_0  \varepsilon)$.

Numerical results of the relative conductance defined in Eq. (\ref{RelativeConductance}) are shown in Fig. 4 (c). By defining the relative conductance, the thermal effect is minimized and the quantum effect is manifested. It can be seen from the figure that variation of the $G_r$ follows the pattern of that of the HM spiral ellipticity. For temperatures above $T_C$ and below $T_N$, $\rm {TbMnO_3}$ is ferromagnetic with its magnetic polarization in the $b/y$ direction. For temperatures below $T_C$, the conductance is suppressed by the HM diffraction. Although at some incident angles the transmission is enhanced by the HM diffraction, it is more strongly suppressed at others. As their combined effect, the angularly-averaged conductance is smaller in the HM phase than in the ferromagnetic phase. Also affected by the ferroelectric polarization, the conductance is larger for decreasing electric potential than increasing electric potential.

It should be noted that we consider an HM-layer with the thickness $d=3.2$ nm, which consists of thirty-one semi-atomic layer in the $c/z$-direction. The direction of the HM spin-field is opposite between two neighboring semi-atomic layers, which is illustrated in Fig. 3. The travelling beam is diffracted once more during the scattering by each semi-atomic layer. Our numerical results demonstrates that the effect of the diffraction over diffraction is simple and not divergent or chaotic. This property at least to some extent justifies our theory and numerical techniques. Also the physics underlying the conductance shown in Fig. 4 and the total transmission shown in Fig. 5 is clear and meaningful.

Properties of the conductance can be illustrated by the total transmission probabilities shown in Fig. 5. The transmission demonstrates diffraction features, i.e., maximums and minimums occur at certain incident angles. Variation of the transmission probabilities as a function of the incident azimuthal angle $\phi _{in}$ differs dramatically for different polar angles $\theta _{in}$. For $\theta _{in} =0$, the transmission is constant for all $\phi _{in}$ due to the symmetry of the HM configuration. The transmission is also symmetric between $\phi _{in} < \pi$ and $\phi _{in} > \pi$ for all $\theta _{in}$. The transmission is largest in normal incidence of $\theta _{in}=0$ and smallest in nearly grazing incidence of $\theta _{in} =1.5$ radian for all incident azimuthal angles $\phi _{in}$. As a result of the diffraction by the HM spiral, maximums and minimums occur. The angularly averaged conductance is suppressed when the total contribution from wavy angular distribution of the transmission is smaller than a constant transmission.

Numerical results of the electrical current, shot noise, and Fano factor at zero temperature are shown in Fig. 6. It is natural that both the electrical current and the shot noise increase with the bias voltage because of more energy channels contributing to the transport. The current and shot noise for different HM spiral helicity characterized by $P_c >0$ and $P_c <0$ are nearly the same. However, the Fano factor decreases with the bias voltage and the two HM helical states are distinctly separated in the Fano factor. The Fano factor indicates the coherent correlation between transport channels. The Fano factor is larger for uncorrelated single-channel transport and smaller for correlated multi-channel transport varying in the range between 0 and 1 for ideal quantum coherent tunneling\cite{BlanterPR2000}. The quantum correlation between different diffraction orders during transport further suppresses the shot noise in addition to the Pauli exclusion and interference in the orbital and spin degrees of freedom compared with the Poisson noise. The Fano factor decreases with the bias voltage because of the strengthening of the correlation between different orbital channels when larger $V$ opens up a wider tunneling window in energy. The difference in the Fano factor between the $P_c>0$ and $P_c<0$ states is a direct effect of the ferroelectricity. The total transmission probability for $P_c>0$ is larger than that for $P_c<0$, which is shown in Fig. 5. As a result, interference is stronger for $P_c>0$ than $P_c<0$, giving rise to a smaller Fano factor.

Numerical results of the electrical current, noise, and the Fano factor as a function of the temperature at a fixed bias voltage of 35 mV are shown in Fig. 7. As an effect of the finite temperature, the noise is made up of the shot noise and the thermal noise. Variation of the finite-bias-voltage electrical current $I$ as a function of the temperature follows a similar pattern to the conductance $G_r$. The properties of the electrical current $I$ are governed by the quantum effect of the scattering by the HM-barrier. When the temperature is decreased, the HM spiral ellipticity changes and the electrical current changes with it. For temperatures above $T_C$ and below $T_N$, $\rm {TbMnO_3}$ is ferromagnetic polarized in the $b/y$ direction. At this temperature range, the ferroelectric polarization is absent and no diffraction occurs. Therefore the current does not change with the temperature and the two cases of $P_c>0$ and $P_c<0$ are not separated. For temperatures below $T_C$, the current is suppressed by the HM diffraction and the two cases of $P_c>0$ and $P_c<0$ are differentiated solely by the electric effect.
In Fig. 7 (b), numerical results of the noise are provided. It can be seen that in the finite-temperature case the thermal effect in the noise is strong and over the quantum effect. The thermal effect could be partially suppressed by removing the thermal noise $S_t$ in defining the Fano factor. The Fano factor in Fig. 7 (c) is much larger than 1 due to contributions made by the Fermi distribution function numerically. The difference between the two cases of $P_c>0$ and $P_c<0$ in the noise and Fano factor is a combined effect of the ferroelectric polarization and the diffraction by the HM-barrier.

\section{Conclusions}

In conclusion, we theoretically investigated the conductance and shot noise properties in the HM tunnel junction. We separately considered a single-HM-layer toy model and a multi-HM-layer real model based on the experimental data of the $\rm{TbMnO_3}$ multiferroic HM material. With the developed scattering matrix scheme, the general procedure and formulas to calculate the diffracted transmission probabilities, the conductance, and the shot noise are given. Numerical results of the single-HM-layer toy model show that at a certain incident angle, one of the diffracted waves $T_1$ or $T_{-1}$ disappears into an evanescent mode and that transmission coefficient abruptly falls into zero giving rise to prominent rise-and-fall oscillations in the angle-averaged shot noise. Two dividing lines of $q = k_F$ and $q= 2 k_F$ characterize the regimes that one or both of the diffracted waves complete disappear into evanescent modes for all incident angles. When $q> 2 k_F$, the shot noise properties resemble that of a plain $\delta$-barrier. At the two sides of $q= k_F$, strength of the two diffracted channel matches each other giving rise to strong oscillation in the shot noise. Due to correlation among different diffracted channels, the shot noise are additionally suppressed relative to the Poisson value. The diffraction-affected transport properties are prominently demonstrated in the shot noise and Fano factor. Numerical results of the multi-HM-layer real model of $\rm{TbMnO_3}$ show that clockwise and counterclockwise spin helix is distinctly separated in the conductance and shot noise spectrum. Variation of the electrical conductance and the finite-bias current as a function of the temperature resembles the variation pattern of the HM spiral ellipticity. The shot noise is further suppressed by the quantum correlation between different diffraction orders during transport in addition to the Pauli exclusion and interference in the orbital and spin degrees of freedom compared with the Poisson noise. The current and shot noise properties are a combined result of helimagnetism and electric polarization.

\section{Acknowledgements}

This project was supported by the National Natural Science
Foundation of China (No. 11004063) and the Fundamental Research
Funds for the Central Universities, SCUT (No. 2014ZG0044).

\clearpage

\begin{figure}[h]
\includegraphics[height=10cm, width=14cm]{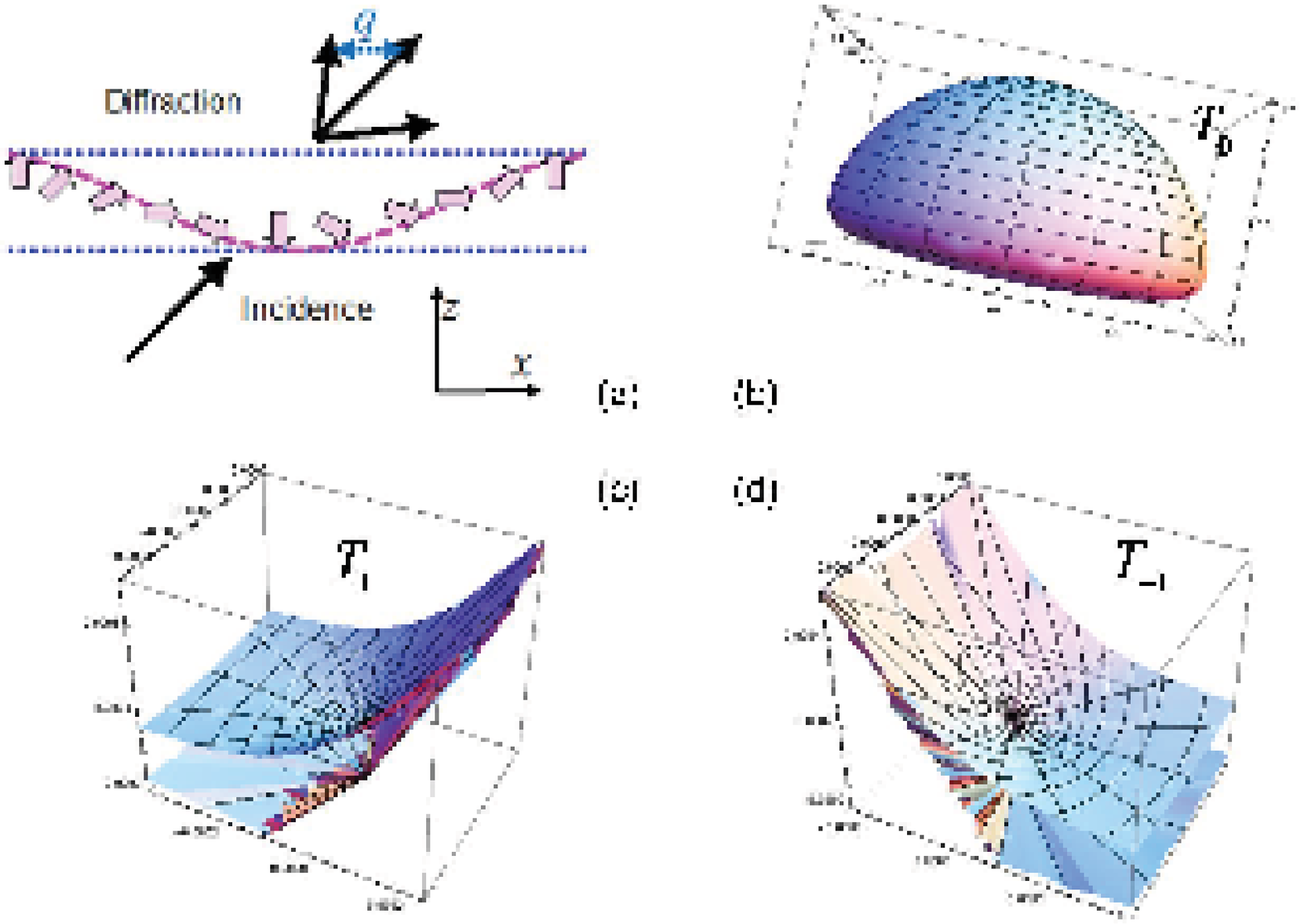}
\caption{(a) Schematic illustration of the diffraction effect in the helimagnet tunnel junction. The helimagnet spin spirals in the $x$-$z$ plane. An incident plane wave can be diffracted into sidebands with the $x$-component wave vector adding or subtracting a spiral wave vector $q$. Higher order diffraction decays exponentially justifying the $\pm 1$ order cutoff. For $q$ larger than one Fermi wave vector $k_F$, one diffraction beam degrades into evanescent surface mode and does not contribute to the transmission. For $q$ larger than $2 k_F$, both diffraction beams degrade into evanescent surface modes and do not contribute to the transmission, in which case the transmission resembles that of a plain $\delta$-barrier. (b) Order-$0$, (c) order-$1$, (d) order-$-1$ transmission $T_0$, $T_1$, $T_{-1}$ in the incident angle space. For panels (b), (c), and (d), $E_F=100$ meV, $V_0=50$ meV$ \cdot $\AA, $J=20$ meV$ \cdot $\AA, $q=0.5 k_F$. Order-$0$ transmission has spherical symmetry. Order-$1$ and $+1$ transmission disappears at certain positive and negative incident angles, respectively. The two surfaces in panels (c) and (d) correspond to the two peaks at which the diffracted beam disappears. One is for grazing incidence, the other is at a certain positive and negative incident angle for $T_1$ and $T_{-1}$, respectively.}
\end{figure}

\clearpage

\begin{figure}[h]
\includegraphics[height=10cm, width=14cm]{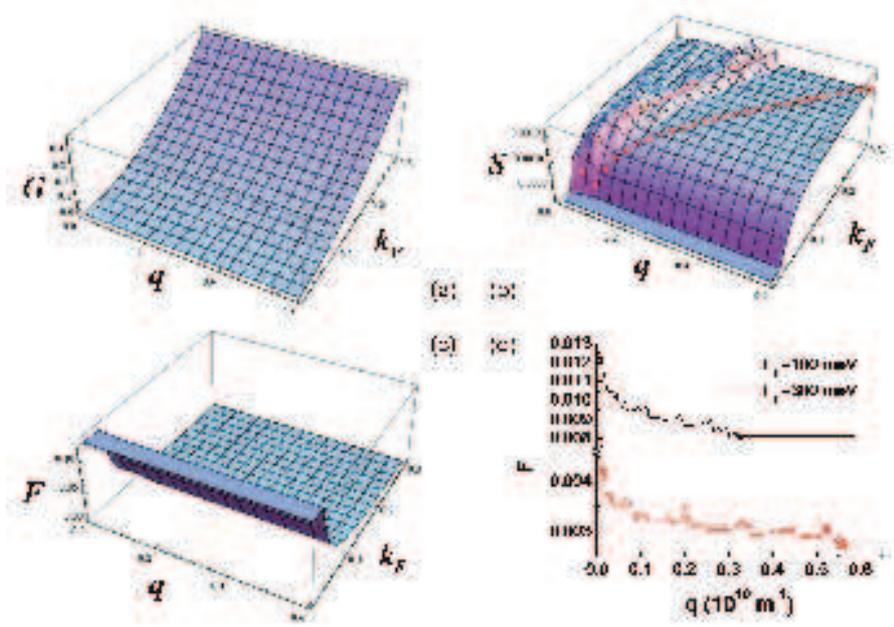}
\caption{Conductance $G$ (a), shot noise $S$ (b), and the Fano factor $F$ (c) as functions of the HM spiral wave vector $q$ and the electron Fermi wave vector $k_F$. $G$ is in unit of $e^2/(h{\rm{\AA}}^2)$. $S$ is in unit of $2e^3/(h{\rm{\AA}}^2)$. In panel (b), the red and pink dotted line corresponds to $q= 2 k_F$ and $q=k_F$, respectively. (d) Fano factor as a function of $q$ for two different Fermi energies. The star symbol is at the Fermi energy with $q=2 k_F$.  }
\end{figure}

\begin{figure}[h]
\includegraphics[height=10cm, width=9cm]{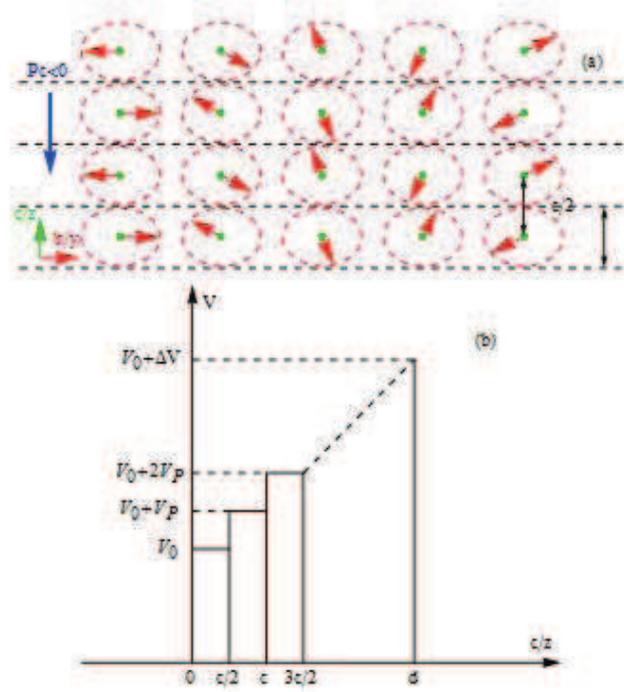}
\caption{ (a) Schematic magnetic structure\cite{YamasakiPRL2007} of the multiferroic helimagnet $\rm{TbMnO_3}$ projected onto the $bc$ plane with $a \times b \times c/2$ defined by the $Pbnm$
orthorhombic cell in the ferroelectric phase below $T_C$=27 K. We use the lattice constants $b=3.9$ \AA and $c=4$ \AA. Translation invariance holds in the $a$ direction including both the $a=0$ and the $a=1/2$ planes. The counterclockwise spiral magnetic structure
corresponds to the electric polarization $P_{c} < 0$, which is sketched in the figure. The clockwise spiral magnetic structure
corresponds to the electric polarization $P_{c} > 0$. We considered both cases in our numerical treatment. In algebra we use the $x$, $y$, and $z$ coordinates to be the $a$, $b$, and $c$ crystal direction, respectively. The crystal was cut into a
thin plate with the widest face of (001) and a thickness of
$d=$3.2 nm, which assumes the $a/x$ and $b/y$ direction infinite\cite{YamasakiPRL2007}. Electrodes are applied to the (001) faces and charges transport in the $c/z$ direction. The diagram draws the beginning four $c/2$ half layers. Lattice constants in the $b$ and $c$ crystal direction used in numerical treatment are 3.9 and 4 \AA, respectively. (b) Energy profile induced by the electric polarization. As a result of the ferroelectric polarization, charges accumulate at the two interfaces of the electrodes connecting the helimagnet sample. The screening charge gives rise to a potential increase of $\Delta U$. Our theoretic model treated the scattering process layer by layer and $\Delta U$ is approximated by small potential steps with each jump of $U_P$.    }
\end{figure}

\begin{figure}[h]
\includegraphics[height=10cm, width=12cm]{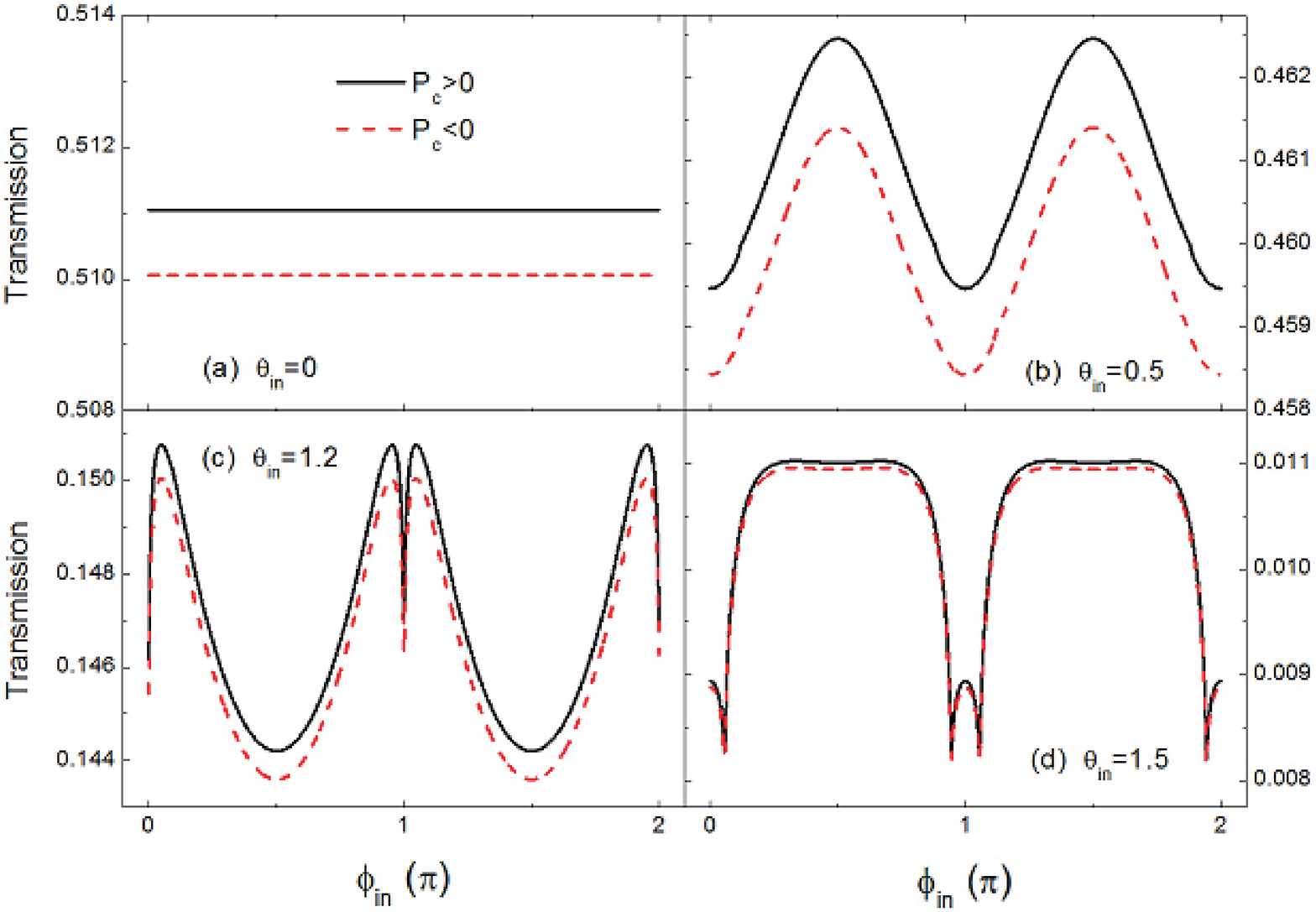}
\caption{ (a) Numerically simulated electric polarization along the $c/z$ axis $P_c$ as a function of the temperature for the multiferroic helimagnet $\rm{TbMnO_3}$. The electric polarization is reversed when the chirality of the helimagnet spiral is reversed from clockwise to counterclockwise\cite{YamasakiPRL2007}. (b) Numerically simulated spiral ellipticity defined as $m_c/m_b$. The KNB model
predicts that the ferroelectric polarization $P_c$ should be proportional
to $m_b m_c$ with $P_c = A \sin (2\pi qb) m_b m_c$. $b$ is the lattice constance along the crystal $b$ direction and $A=40$ $\mu $F is an experimentally determined constant\cite{YamasakiPRL2007}. (c) Relative conductance $G_r$ defined by Eq. (\ref{RelativeConductance}) as a function of the temperature. }
\end{figure}

\begin{figure}[h]
\includegraphics[height=10cm, width=13cm]{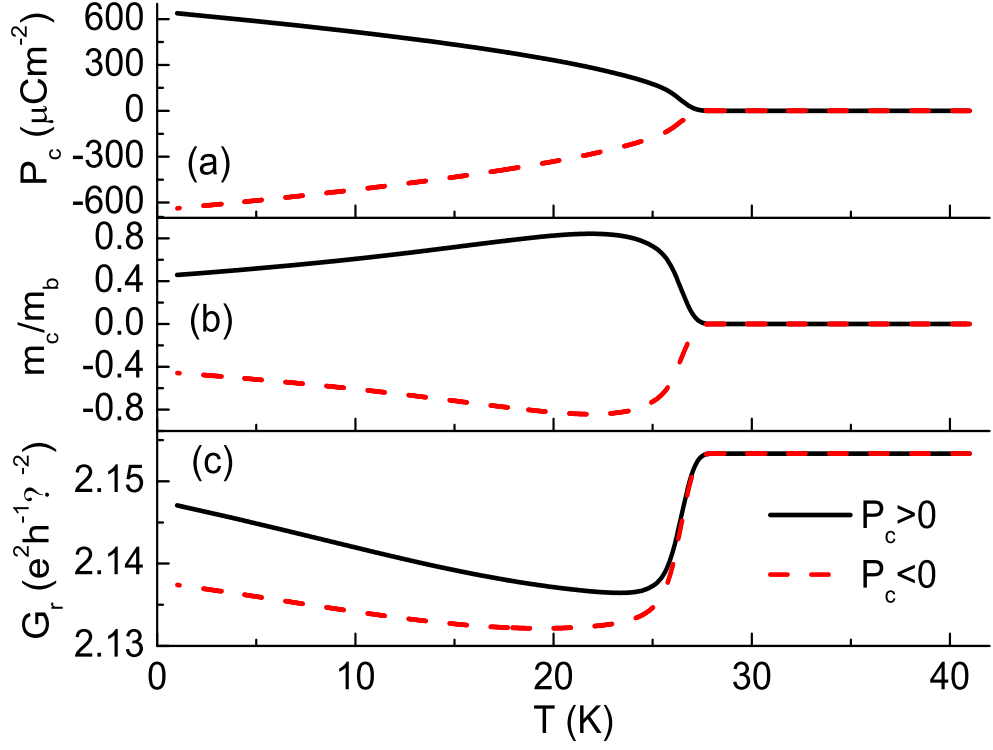}
\caption{ Total transmission probability at $T=15$ K as a function of the incident azimuthal angle $\phi _{in}$ for different incident polar angles $\theta _{in}$. At this temperature\cite{YamasakiPRL2007} $m_b=3.9$, $m_c=2.8$, and $P_c=433$ $\mu $F. The black solid and red dashed lines correspond to the two HM chiral states with $P_c>0$ and $P_c<0$, respectively. $\phi _{in}$ is measured in $\pi$ and $\theta _{in}$ is measured in radian.   }
\end{figure}

\begin{figure}[h]
\includegraphics[height=10cm, width=13cm]{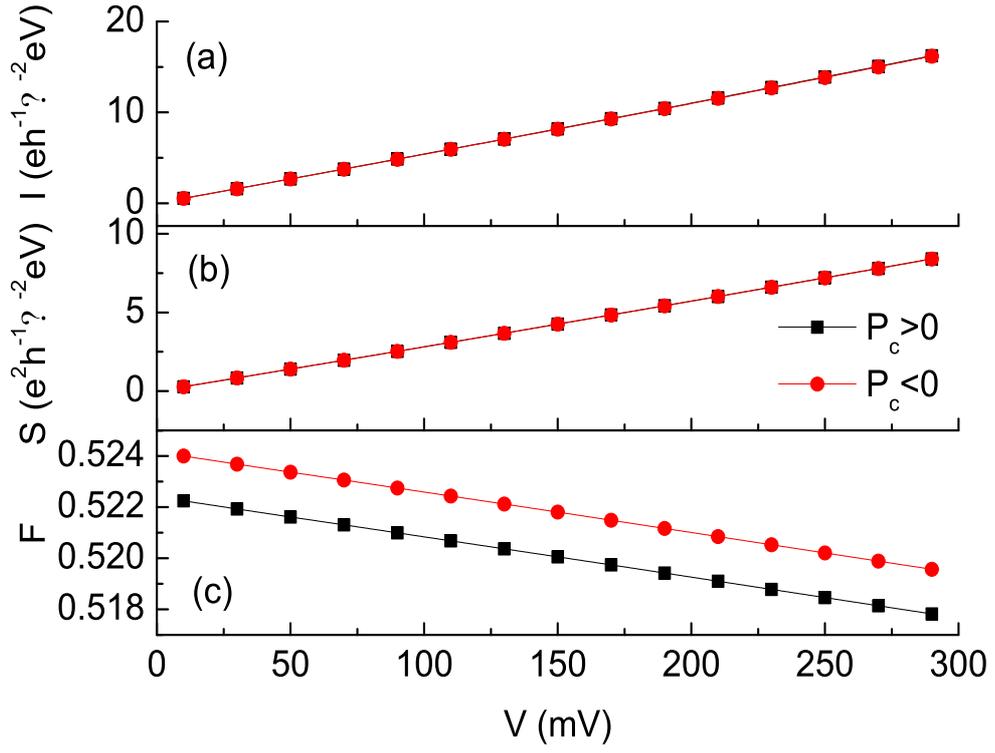}
\caption{ Electrical current, shot noise, and the Fano factor as a function of the bias voltage at zero temperature for different HM spiral helicity characterized by $P_c>0$ and $P_c<0$. Helimagnetic and ferroelectric parameters are taken from Fig. 4. }
\end{figure}

\begin{figure}[h]
\includegraphics[height=10cm, width=13cm]{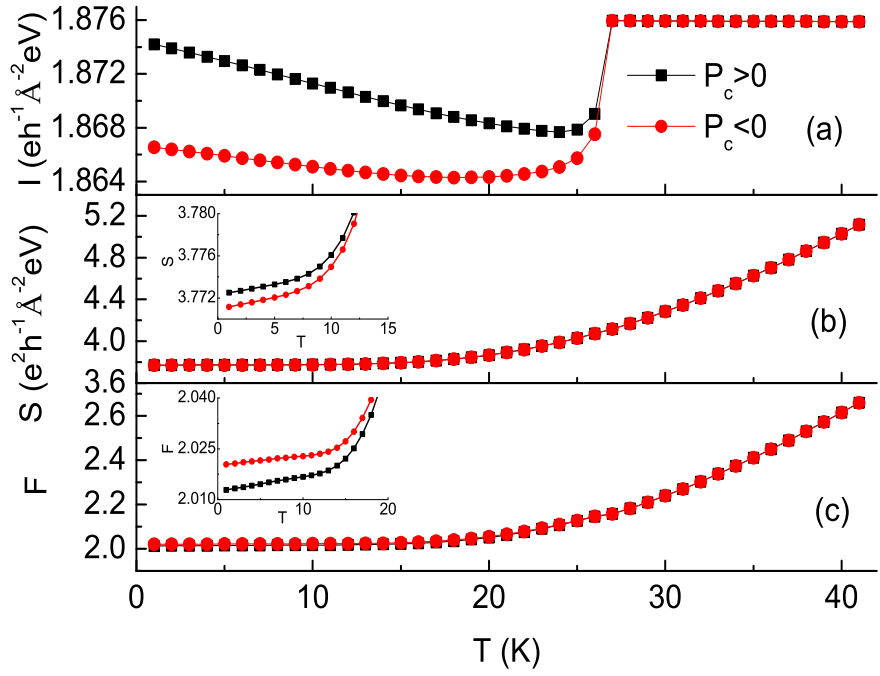}
\caption{ Electrical current, noise, and the Fano factor as a function of the temperature for different HM spiral helicity characterized by $P_c>0$ and $P_c<0$. As an effect of the finite temperature, the noise is made up of the shot noise and the thermal noise. Insets in panels (b) and (c) are zoom-in of the low temperature region to promote the difference between the two curves. Helimagnetic and ferroelectric parameters are taken from Fig. 4. The bias voltage $V$ is fixed to be 35 mV. }
\end{figure}

\clearpage

\end{document}